\documentstyle[12pt,aps,epsfig]{revtex}

\def\pmb#1{\mbox {\boldmath $#1$}}

\begin{document}
\title{Screened electrostatic interactions between clay platelets.}
\author{D.G. Rowan$^*$, J-P. Hansen$^*$ \& E.Trizac$^{**}$}
\maketitle
\begin{center}
* Department of Chemistry, University of Cambridge, Lensfield Road, Cambridge CB2 1EW, United Kingdom\\ 
*** Laboratoire de Physique Th\'eorique (UMR 8627), B\^at. 211, Universit\'e Paris-Sud, 91405 Orsay Cedex, France
\end{center}

\begin{abstract}
An effective pair potential for systems of uniformly charged
lamellar colloids in the presence of an electrolytic solution of 
microscopic  co- and counterions is derived.  The charge distribution on the 
discs is expressed as a collection of multipole moments, and the 
tensors which determine the interactions between
these multipoles are derived from a screened Coulomb potential.  
Unlike previous theoretical studies of such systems, the interaction
energy may now be expressed for discs at arbitrary mutual orientation.
The potential is shown to be exactly equivalent to the
use of linearized Poisson-Boltzmann theory.
\end{abstract}
\section{Introduction}
While the mesostructure, stability and phase behaviour of charge-stabilized
dispersions of spherical colloidal particles are by now reasonably well 
understood, both experimentally and theoretically \cite{verwey,arora,vanroij}, the
picture is much less clear in the case of lamellar colloids, of which
clay dispersions are a pre-eminent example \cite{vanolphen}.  This is partly
due to the high degree of polydispersity, the irregular shapes, and the extreme
anisotropy of the thin lamellar particles of naturally occurring clay suspensions.   Such complications render an unambiguous interpretation of experimental
data, eg. from small angle X-ray or neutron diffraction measurements, very
difficult, while posing a practically insurmountable challenge to the theoretician attempting  a Statistical Mechanics description.  Even for the widely
studied synthetic model system of Laponite, made up of nearly monodisperse, disc-shaped platelets, there is no consensus among experimentalists as to the 
structure, gelling behaviour and rheology of semi-dilute suspensions
\cite{ramsay,mourchid,pignon,kroon,bonn}, while attempts at a theoretical description, or simulations
of this model system are in their infancy \cite{dijkstra,kutter}. The main reason
for the latter state of affairs is that a realistic model for the effective
interaction between a pair of arbitrarily oriented charged circular platelets,
generalizing the isotropic DLVO potential \cite{verwey} between spherical
colloids, is not available. Only in the simplest case of two coaxial,
uniformly charged discs, has a screened Coulomb interaction been worked out within linear \cite{trizac1} and non-linear \cite{carvalho} Poisson-Boltzmann (PB) theory.  

The Molecular Dynamics (MD) simulations of ref.[11] were based on a 
site-site interaction model, generalizing the ``Yukawa segment'' representation used earlier to simulate suspensions of charged rods \cite{lowen1}. Within such
a representation the charge distribution on a rod or a platelet is discretized
into $\nu$ interaction sites, each carrying a fraction $1/\nu$ of the total
charge; sites on different particles interact via  a screened Coulomb potential. This Yukawa segment model is very computationally intensive, since the total
interaction between two particles involves $\nu^2$ contributions.  It also
carries a degree of arbitrariness in the choice of the number of sites $\nu$, which for practical reasons must be taken to be much smaller than the number of elementary surface charges (typically $10^3$ for Laponite) carried by
individual platelets.  

This paper examines the continuous version of the Yukawa segment model, and 
derives a multipolar expansion of the effective, screened Coulomb interaction between two uniformly charged platelets of arbitrary relative orientations. The resulting anisotropic effective pair potential is shown to be accurate for 
centre-to-centre  distances larger than the radius of the platelets, and should
hence provide a useful tool for theoretical investigations of the structure and
sol-gel transition in semi-dilute clay dispersions.

\section{The Yukawa segment model}
\label{sec:introduction}

Consider a suspension of $N_p$ infinitely thin circular platelets per unit volume, of radius $a$, and carrying a uniform surface charge density $\sigma=Ze/\pi a^2$, where $Ze$ ($< 0$) is the total charge on a platelet. The platelets, 
together with microscopic co- and counterions, are suspended in water.  Since the present study focuses  on mesoscopic lengthscales, of the order of $a$ ( typically $a\simeq 15\rm{nm}$ for Laponite), one may neglect the molecular nature of water which will be regarded as a continuum of dielectric constant $\epsilon$. Within linearized PB theory, the effective interactions between platelets are always pairwise additive \cite{lowen2}, and the screening of electrostatic interactions by the co- and counterions is uniquely characterized by the Debye
screening length:
\begin{equation}
\label{debyescreeninglength}
\lambda_D=\frac{1}{\kappa}=\left(\sum_\alpha \frac{n_\alpha z_\alpha^2 e^2}{\epsilon_0\epsilon k_B T}\right)^{-\frac{1}{2}}
\end{equation}
where the sum is over all microion species, $n_\alpha$ and $z_\alpha$ are the concentration (number density) and valence of ions of species $\alpha$, and 
$\epsilon_0$ is the permittivity of free space.  Building on the linearity of the theory, the Yukawa segment model assumes that each infinitesimal area $ds$ 
(or ``segment'') on a uniformly charged disc generates a screened Coulomb 
potential:
\begin{equation}
\label{elementpot}
\phi(r)=\frac{\sigma ds}{4\pi\epsilon_0 \epsilon r}e^{-\kappa r}.
\end{equation}
The corresponding pair potential between two infinitesimal areas $ds$ (around ${\pmb r}$) and $ds^{'}$ (around ${\pmb r}^{'}$) on two discs is then
\begin{equation}
\label{elementpair}
v(|{\pmb r}-{\pmb r}^{'}|)= \frac{\sigma^2 ds ds^{'}}{4\pi\epsilon_0\epsilon |{\pmb r}-{\pmb r}^{'}|}e^{-\kappa|{\pmb r}-{\pmb r}^{'}|},
\end{equation}
and the total pair interaction between the discs is obtained by integrating 
(\ref{elementpair}) over the surfaces of the two discs. However, for
arbitrary orientations of the discs, this leads to intractable expressions
involving multiple integrals.

Instead, by analogy with electrostatic interactions between extended charge distributions,  a systematic multipolar expansion of the screened Coulombic
interaction will be sought. The derivation of such an expansion requires some care,
because the basic screened Coulomb potential (\ref{elementpot}) does not satisfy Poisson's equation, except in the bare Coulomb limit, where $\kappa\rightarrow 0$.

\section{The potential around a single plate}

To get a feeling for a  multipolar expansion involving screened, rather than
bare, Coulomb interactions, consider first the potential due to a uniformly
charged disc, along the axis of the disc.  Using cylindrical coordinates, with
the $z-$coordinate along the axis of the disc (cf. Fig. \ref{figure1}), the screened 
potential along that axis (ie. for a radial coordinate $\rho =0$) is
simply 
\begin{eqnarray}
\label{barepot1}
\Psi(\rho=0, z)&=&\frac{2\pi\sigma}{4\pi\epsilon_0\epsilon }\int_0^a 
\frac{e^{-\kappa\sqrt{R^2+z^2}}}{\sqrt{z^2+R^2}} RdR\nonumber\\ 
&=& \frac{\sigma e^{-\kappa|z|}}{2\epsilon_0\epsilon \kappa}[1-e^{-\kappa|z|[\sqrt{1+\left(\frac{a}{z}\right)^2}-1]}]
\end{eqnarray}
which is easily expanded in powers of $(a/z)$ according to
\begin{equation}
\label{screen-z}
\Psi(\rho=0,z)=\frac{Ze}{4\pi\epsilon_0\epsilon} e^{-\kappa|z|} \sum_{n=0}^{\infty} \left[A_n +K_n\right]\left(\frac{1}{|z|}\right)^{n+1},
\end{equation}
where the coefficients $A_n$ and $K_n$ are listed in Table \ref{table1}.

Several points are to be noted about this expansion. First, the corresponding expansion for the bare Coulomb potential is correctly retrieved by taking the limit $\kappa\rightarrow 0$; this amounts to setting all $K_n=0$ in 
eq.(\ref{screen-z}), leaving only odd powers of $1/z$ in the expansion, since all 
odd coefficients $A_n$ vanish.  This is an obvious consequence of the space reflection symmetry of the uniform charge distribution on a disc, which implies that
only even multipole moments exist.  However, for the screened Coulomb potential,
terms with even powers of $1/z$ appear in the expansion,
which would correspond to odd multipoles (dipole etc.) in the bare Coulomb case. 

The second remark is that the expansion (\ref{screen-z}) also follows
from the exact potential due to a uniformly charged disc, and its associated electric double-layer of co- and counter ions, as calculated within linearized
PB theory \cite{ref16}, namely:
\begin{equation}
\label{philinearizedpb}
\Psi(\rho,z)=\frac{2Ze}{a\epsilon_0\epsilon}\int_0^\infty J_1(ka)J_0(k\rho) \frac{e^{-|z|\sqrt{k^2+\kappa^2}}}{\sqrt{k^2+\kappa^2}}dk.
\end{equation}
This agreement between the expansion in eq.(\ref{screen-z}) and the expansion of
eq.(\ref{philinearizedpb}) for $\rho=0$ is an illustration of the exact equivalence between the Yukawa segment model, and a full linearized PB calculation
of the effective potential generated by a charged particle of any shape and its associated electric double layer.

The final point concerns the generalization of the expansion (\ref{screen-z}) away from the $z-$axis, ie for $\rho \ne 0$.  In the bare Coulomb case $(\kappa=0)$, the coefficients $A_n$ in table 1 may be immediately carried over to 
spherical polar coordinates $(r,\theta,\phi)$ to write down an expansion of
the potential due to a uniformly charged disc in even Legendre polynomials (the potential is independent of the azimuthal angle $\phi$):
\begin{equation}
\label{screenlegend}
\Psi(r,\theta)=\frac{Ze}{4\pi\epsilon_0\epsilon}\sum_{n=0;\ \rm{even}}^\infty A_n \frac{1}{r^{n+1}} P_n(\cos\theta).
\end{equation}
However, the presence of the exponential screening factor in the expansion (\ref{screen-z}) prevents a similar straightforward generalization to off-axis conditions in
the screened Coulomb case ($\kappa\neq0$).  For this reason the multipolar expansion must be re-examined more carefully for the Yukawa segment model.

\section{Screened multipolar expansion}

The multipolar expansion of the total potential $\Psi({\pmb r})$ due to a uniformly charged disc, with each infinitesimal surface element generating the screened potential (2) (Yukawa segment model), may be derived along lines
nearly identical to the classic calculation for unscreened charge
distributions \cite{jackson}. Clearly 
\begin{equation}
\label{Psir}
\Psi({\pmb r}) = \int_S \phi(|{\pmb r}-{\pmb s}|) d{\pmb s}
\end{equation}
where the integral is over the surface $S$ of the disc. The potential 
$\phi$ may now be expanded as a Taylor series about the centre of the disc $({\pmb s}={\pmb 0})$, ie.:
\begin{equation}
\label{explainpot1}
\Psi({\pmb r})=\sigma \int_S d{\pmb s} \left\{ \phi (r) -\sum_\alpha s_\alpha \nabla_\alpha \phi(r) + \frac{1}{2!} \sum_{\alpha\beta}s_\alpha s_\beta \nabla_\alpha \nabla_\beta \phi(r) +\cdots \right\},
\end{equation}
where the sums are over all three Cartesian coordinates of the vector ${\pmb s}\in S$.  All odd terms (eg. the dipolar term) vanish by symmetry. This leaves only the even terms:
\begin{equation}
\label{PsiR2}
\Psi({\pmb r})=\Psi_Z(r)+\Psi_Q({\pmb r}) + \Psi_\Phi({\pmb r}) +\dots,
\end{equation}
involving the total charge $Ze$ of the disc, its quadrupole tensor $\underline{\underline{\pmb Q}}$, its hexadecapole tensor ${\underline{\underline{\pmb \Phi}}}$, etc.:
\begin{eqnarray}
\Psi_Z(r)&=&ZeT^\kappa(r)\nonumber\\
\label{equation11b}
\Psi_Q({\pmb r})&=&\frac{e}{2!}Q_{\alpha\beta}T^{\kappa}_{\alpha\beta}({\pmb r})\\
\Psi_\Phi({\pmb r}) &=& \frac{e}{4!}\Phi_{\alpha\beta\gamma\delta}T^{\kappa}_{\alpha\beta\gamma\delta}({\pmb r})\nonumber
\end{eqnarray}
where the Einstein convention of summation over repeated indices has
been adopted.  The tensors $T^\kappa$ are:
\begin{equation}
\label{ktensor}
T_{\alpha\beta\gamma\cdots}^\kappa = \nabla_\alpha \nabla_\beta \nabla_\gamma \cdots \left(\frac{1}{4\pi\epsilon_0 \epsilon}\frac{e^{-\kappa r}}{r}\right)
\end{equation}
while the $Q_{\alpha\beta},\Phi_{\alpha\beta\gamma\delta}$ are the Cartesian
components of the $2^{nd}$ rank quadrupolar and $4^{th}$ rank hexadecapolar
tensors.  For a uniformly charged disc the quadrupolar tensor is given (in 
a frame where the $z$-coordinate is along the axis of the disc) by
\begin{eqnarray}
\label{defineq}
\underline{\underline{{\pmb Q}}}&=&\frac{\sigma}{e}\int_S   {\pmb s}{\pmb s} d{\pmb s}\nonumber\\
&=& \left(\begin{array}{ccc}
-Q & 0 & 0 \\
0 & -Q & 0 \\
0 & 0 & 0 
\end{array}\right)
\end{eqnarray}  
where $Q=-Za^2/4$.  Note that contrary to the bare Coulomb case, $\underline{\underline{\pmb Q}}$ cannot be chosen traceless in the screened case, because
the tensors $T^\kappa_{\alpha\beta\cdots}$ are themselves not traceless.  The Cartesian components of the $4^{th}$ rank hexadecapole moment are defined by:
\begin{equation}
\label{hexcomp}
\Phi_{\alpha\beta\gamma\delta}=\frac{\sigma}{e}\int_S s_\alpha s_\beta s_\gamma s_\delta d{\pmb s}. 
\end{equation}
For a disc, choosing the $z-$coordinate along its axis, the only non-zero components  $\alpha\beta\gamma\delta$ are the two diagonal components 
$\Phi_{xxxx}$ and $\Phi_{yyyy}$ and those 
in which $x$ and $y$ both appear twice. Explicitly
\begin{eqnarray}
\Phi_{xxxx}=\Phi_{yyyy} = \frac{Za^4}{8} &\equiv& \Phi\nonumber\\
\label{phidefined}
\Phi_{xxyy}=\Phi_{xyxy} = \cdots = \frac{Za^4}{24}&\equiv& \frac{\Phi}{3}.
\end{eqnarray}
The calculation of the tensors $T_{\alpha\beta\gamma\dots}^{\kappa}$ is considerably lengthier for the screened than for the bare Coulomb interaction. Some details are given in Appendix A. 

In spherical coordinates, the total potential due to the uniformly charged disc, up to hexadecapolar order, may finally be written as:
\begin{eqnarray}
\label{Psifull}
\Psi(r,\theta)&=& \frac{e}{4\pi\epsilon_0\epsilon}e^{-\kappa r} \left\{
\frac{1}{r}\left[Z+\frac{Q}{2}\kappa^2(\cos^2\theta-1) +\frac{\Phi}{24}\kappa^4\sin^4\theta\right]\right. \nonumber\\
&\ & \hspace{1.6cm} +\frac{1}{r^2}\left[\frac{Q}{2}\kappa(3\cos^2\theta-1)-\frac{\Phi}{3}\kappa^3\left(\sin^2\theta - \frac{5}{32}\sin^4\theta\right)\right] \nonumber\\
&\ & \hspace{1.6cm} + \frac{1}{r^3}\left[ \frac{Q}{2}(3\cos^2\theta -1)+\frac{\Phi}{3}\kappa^2\left(1- 6\sin^2\theta +  \frac{45}{8}\sin^4\theta\right)  \right]\nonumber\\
&\ & \hspace{1.6cm} + \frac{1}{r^4}\left[\frac{\Phi}{8}\kappa(35\cos^4\theta-30\cos^2\theta+3) \right]\nonumber\\
&\ & \hspace{1.55cm} \left.+
\frac{1}{r^5}\left[\frac{\Phi}{8}(35\cos^4\theta-30\cos^2\theta+ 3)  \right]\right\}
\end{eqnarray}            
Returning to cylindrical coordinates, this expression reduces on the $z-$axis
to:
\begin{eqnarray}
\label{Psizaxis}
\Psi(\rho=0,z)&=& \frac{Ze}{4\pi\epsilon_0\epsilon}e^{-\kappa|z|}\left[\frac{1}{z}-\frac{\kappa a^2}{4}\left(\frac{1}{z}\right)^2 + \left(\frac{\kappa^2a^4}{24} -\frac{a^2}{4}\right)\left(\frac{1}{z}\right)^3\right.\nonumber\\
&\ & \hspace{2.0cm} + \left.\frac{\kappa a^4}{8}\left(\frac{1}{z}\right)^4 + \frac{a^4}{8}\left(\frac{1}{z}\right)^5\right]
\end{eqnarray}
which coincides with the expansion (\ref{screen-z}), with coefficients given in
Table 1, up to order $a^4$ (the higher powers of $a$ in the coefficients of
$1/z^4$ and $1/z^5$ corresponding to higher order multipole moments).
To illustrate the convergence of the multipolar expansion of the potential to hexadecapolar order, in figs. \ref{figure2} and \ref{figure3} the 
potential given by (\ref{Psifull}) is compared with the linearized PB potential of eq.(\ref{philinearizedpb}) and an explicit numerical integration over a discretized charge distribution (with $\nu=7841$ sites), both along the $z-$axis and in the $xy-$plane. The numerical integration was carried out to
check its accuracy against the exact potential (\ref{philinearizedpb}), since it
will be the only available test of the multipolar expansion of the screened pair
interaction between two platelets at arbitrary orientations, to 
be discussed in section VI. 
Agreement is seen to be excellent, down to around $z(\rho)\simeq a$, where
the multipolar expansion diverges dramatically, as higher order terms in $(1/r)^n$ start to dominate.

\section{Effective  interactions between two plates}
The results of the previous section for the effective potential around a 
single platelet may now be used to determine the potential energy of interaction (or effective pair potential) of a pair of discs, at arbitrary relative
orientations, as shown in Fig.\ref{figure4}. The interaction energy 
$V_{AB}$ is formally expressed by integrating the screened electrostatic potential arising from one disc $(A)$ defined by eqs.(\ref{PsiR2}) and (\ref{equation11b}), over
the surface charge distribution of the second disc $(B)$:
\begin{equation}
\label{expint1}
V_{AB}(r,\theta_A,\theta_B,\phi_A,\phi_B)= \int_{S_B} \sigma d{\pmb s} \Psi^A({\pmb r}+{\pmb s}).
\end{equation}
The  potential $\Psi^A({\pmb r}+{\pmb s})$ may be expanded in a Taylor series 
in powers of ${\pmb s}$, about the centre of disc B ({\pmb s}={\pmb 0}), along the lines of eq.(\ref{explainpot1}), with the
result:
\begin{equation}
\label{expint2}
V_{AB} = \sigma \int_{S_B}d{\pmb s} \left\{ \Psi^A({\pmb r}) +s_\alpha^B \nabla_\alpha \Psi^A({\pmb r}) +\frac{1}{2!}s_\alpha^B s_\beta^B 
\nabla_\alpha \nabla_\beta \Psi^A({\pmb r}) + \cdots \right\}.
\end{equation}
Now inserting the expansions of the potential $\Psi^A({\pmb r})$  (\ref{PsiR2}),  and its derivatives,  into (\ref{expint2}) the electrostatic pair potential is
given by
\begin{eqnarray}
\label{expint3}
V_{AB}&=& e\sigma \int_{S_B} d{\pmb s}^B \left\{ \left[T^\kappa Z^A + \frac{1}{2!}T^\kappa_{\alpha\beta}Q_{\alpha\beta}^A+ \frac{1}{4!}T^\kappa _{\alpha\beta\gamma\delta}\Phi_{\alpha\beta\gamma\delta}^A + \cdots \right]\right.\nonumber\\
&\ & \hspace{1.0cm} + s_\alpha\left[T^\kappa_\alpha Z^A +\frac{1}{2!}T^\kappa_{\alpha\beta\gamma}Q_{\beta\gamma}^A + \frac{1}{4!}T^\kappa_{\alpha\beta\gamma\delta\epsilon}Q_{\beta\gamma\delta\epsilon}^A +\cdots \right] \nonumber\\
&\ & \hspace{1.0cm} +\left. \frac{1}{2}s_\alpha s_\beta \left[T^\kappa_{\alpha\beta}Z^A +
\frac{1}{2!}T^\kappa_{\alpha\beta\gamma\delta}Q_{\gamma\delta}^A +\cdots\right] +\cdots\right\}.
\end{eqnarray}
Integrating  over the surface, $S_B$,  of disc B naturally introduces the 
multipole moments of the surface charge distribution of that disc into
the expression (\ref{expint3})  for $V_{AB}$, which may
hence be cast in the form  
\begin{equation}
\label{utottot}
V_{AB}(r,\theta_A,\theta_B,\phi_A,\phi_B)=V_{AB}^{ZZ}+(V_{AB}^{ZQ}+V_{AB}^{QZ}) + V_{AB}^{QQ} +(V_{AB}^{Z\Phi}+V_{AB}^{\Phi Z}) +\cdots
\end{equation}
where
\begin{eqnarray}
\label{equationuaba}
V_{AB}^{ZZ} &=& e^2Z^A T^\kappa Z^A\\
\label{equationuabb}
V_{AB}^{ZQ} &=& \frac{e^2}{2!} Z^A T_{\alpha\beta}^\kappa Q_{\alpha\beta}^B \\
\label{equationuabc}
V_{AB}^{QQ} &=& \frac{e^2}{2!2!} Q_{\alpha\beta}^A T_{\alpha\beta\gamma\delta}^\kappa Q_{\gamma\delta}^B \\
\label{equationuabd}
V_{AB}^{Z\Phi} &=& \frac{e^2}{4!} Z^A T_{\alpha\beta\gamma\delta}^\kappa \Phi_{\alpha\beta\gamma\delta}^B
\end{eqnarray}
This expansion is consistent up to order $1/r^5$ with the corresponding
series for pure Coulombic interactions ($\kappa=0$).
The rather  cumbersome expressions for the screened interaction tensors 
are  given in Appendix A,  while details of the summations over Cartesian 
coordinates, implicit in eqs.(\ref{equationuabb}-\ref{equationuabd}),  
are described in appendix B.

\section{Results for specific geometries}

The detailed behaviour of the effective pair potential $V_{AB}$ defined by
eqs.(\ref{utottot}-\ref{equationuabd}), will now be examined, as a
function of the centre-to-centre distance between, and the mutual orientations
of, the two discs.  The relevant variables are   
\begin{itemize}
\item r, the separation of the centres of masses of the two discs
\item $\{\theta_A,\theta_B\}$ the polar angles of the two discs
\item $\Delta \phi$, the difference in the azimuthal angles of the 2 discs
\item $\kappa$, the inverse screening length determined by the co- and counterions,
\end{itemize}
Rather than specialize to the physical parameters particular to Laponite ($Z\simeq -1000, a\simeq 15$ nm),  in the figures which follow all distances shall be
scaled by the disc radius $a$, which provides a convenient lengthscale,
and all energies by the bare Coulombic
energy of two charges $Ze$ a distance $a$ apart, ie. by $Z^2e^2/4\pi\epsilon_0\epsilon a$. Thus these results apply to the most general uniformly charged disc.
 
The case of two coaxial plates is illustrated in Fig. \ref{figure5}, where
the various contributions to the energy $V_{AB}$ are plotted as a
function of the distance between the two plates.  The exact result in this 
simple geometry is known within LPB theory \cite{trizac1}, and used to test
the convergence of the multipolar series.  As expected, the 
charge-charge contribution (\ref{equationuaba}) dominates for large spacings, but the
contributions of the higher order terms become very significant
for spacings less than the disc diameter, $2a$.

In Fig. \ref{figure6}  the energy is plotted as a function of the separation between
two discs lying in the same plane, like two coins on a table.  In
this geometry all multipolar contributions are positive (repulsive),
and the enhancement of the total energy over the charge-charge term is
more pronounced. Included in this figure is the potential energy
obtained by integrating over the Yukawa potential over two 
discretized site-charge distributions, where each disc has $\nu=7841$ 
point charges, each carrying charge $Ze/\nu$. The agreement of this
calculation with the multipolar expansion is excellent, down to the
distance of closest approach $r=2a$ below which point the discs intersect.

In Fig. \ref{figure7} the energy is plotted as a function 
of the separation of two discs in a
T-shaped configuration ($\theta_A=0,\theta_B=\pi/2$), a geometry  
favoured by the quadrupoles, as also observed in ref.\cite{dijkstra} where a 
purely quadrupolar disc model was studied.  The agreement with the
discretized Yukawa segment calculation is again excellent, except
at very close centre-to-centre separations, where the
multipolar expansion is expected to collapse.  

In the next plot, Fig.\ref{figure8}, the behaviour of the pair potential
is examined as one disc slides over the other at constant altitude ($z=1.5a$ in this plot), with both discs parallel to each other.  As expected the
energy goes through a maximum at the distance of closest approach, but
detailed structure is seen in the total energy, as each order of 
multipole-multipole interaction decays with differing power law 
behaviour.  The agreement with the numerical Yukawa segment calculation
is again good, except in the region around closest approach, where the
platelets are co-axial, and the greatest deviation is observed, as
seen in Fig.\ref{figure5}. This deviation decreases significantly the greater the
vertical separation of the platelets, and may be improved at close
separation by the inclusion of higher order multipole moments.

In Fig. \ref{figure9}, the angular dependence of the potential
is examined by varying the angle $\theta_B$ at fixed centre-to-centre
separation of $r=1.5a$ and $\theta_A=0$. The charge-charge interaction
is obviously independent of relative orientation, but significant variation
in all higher order interactions is observed. Notably the quadrupole-quadrupole 
interaction is favoured when the discs are perpendicular to each other.
The agreement with the computationally expensive numerical integration method
is observed to be good, except when the discs are parallel, where the
deviation is most noticeable, as observed and commented upon in Figs. 5 and 8.

Finally in Fig. \ref{figure10} the dependence of the interaction on
the azimuthal angles is probed, for disc separations in the range $1<r/a<1.5$ for discs fixed at $\theta_A=\theta_B=\pi/4$.
As noted in Appendix B the interaction energy only depends on the
difference $\Delta\phi=\phi_A-\phi_B$. As the separation of the discs
increases the angular dependence of the pair potential is seen to diminish,
and indeed at large separations disappears.
%%
% - insert figure 10 here
%%
\section{Conclusion}
The familiar multipole moment expansion of the electrostatic interaction
between two extended charge distributions have been generalized to the case where the
interaction is linearly screened by co- and counterions of an 
ionic solution in which the charged colloidal particles are
dispersed.   The case of uniformly charged, disc-like platelets, a model for
the synthetic clay Laponite, was specifically considered in this paper, but the Yukawa-segment model, and the corresponding multipolar 
expansion, may be extended to particles of any shape.  For example
this procedure may readily be adapted to the case of uniformly charged rods,
where symmetry again precludes multipole moments of odd order,
by insertion of the relevant quadrupole and hexadecapole tensors for
such a charge distribution.  In the simplest  case of spherical particles, the
present treatment leads back to the familiar DLVO potential.  For discs, 
the expansion, including up to quadrupole-quadrupole and charge-hexadecapole terms, yields interaction energies in good agreement with data for
a discretized version of the Yukawa segment model, down to centre-to-centre
distances of the order of the disc radius $a$, for all relative orientations
of the two platelets which were investigated.  As expected, the expansion
breaks down at shorter distances, and yields rapidly divergent energies as
$r\rightarrow 0$.

The effective pair potential defined by eq.(\ref{utottot}), and the
explicit expressions in the Appendices, should prove useful in Statistical
Mechanics descriptions of semi-dilute clay dispersions, and of their sol-gel
transition, provided suitable short-range cut-offs are imposed.  For
computer simulations of more concentrated dispersions, an appropriate
strategy would be to use a hybrid pair potential approach, interpolating
between the multipolar expansion at large centre-to-centre distances, and
a direct summation of the $\nu^2$ screened Coulomb site-site interactions
in a discretized version of the Yukawa segment model, similar to that used in 
Ref.\cite{kutter}, at short distances.

In order to determine the phase behaviour of dispersions of charged platelets, from direct calculations of the free energy of systems of platelets interacting
via the multipolar effective pair potential derived in this paper, it is important to include a structure-independent `volume' term in the free
energy \cite{grimson}. Such a volume term has been shown to play a crucial
role in the determination of phase diagrams of suspensions of spherical charged
colloidal particles, in the regime of very low concentration of
added electrolyte \cite{vanroij,vr3}.  The exact form of the `volume' term
can be determined from a careful analysis of the density functional
formulation of linearized PB theory \cite{vanroij,lowen2}. Such an
analysis, which also provides a rigorous foundation of the Yukawa segment
model \cite{kutter}, is under way.

\section{Acknowledgments}
The authors would like to thank H. L\"owen and A.J. Stone for useful
discussions, and
gratefully acknowledge the support of the Franco-British Alliance
program, project no. $PN 99.041$.  DGR would like to thank the EPSRC 
for their continued support.

\begin{appendix}
\section{Cartesian Tensors for a screened Coulomb interaction}
The Cartesian tensors for a screened Coulomb interaction, are defined 
(analogously to the bare Coulomb case) as derivatives of the
potential, ie:
\begin{equation}
T_{\alpha\beta\gamma\dots}^{\kappa}=\nabla_\alpha \nabla_\beta \nabla_\gamma \dots \left(\frac{1}{4\pi\epsilon_0\epsilon}\frac{e^{-\kappa r}}{r}\right)
\end{equation}
Now, for any function $f(r)$, ($f(r)$ being $e^{-\kappa r}/r$ in the 
current work) 
the gradient $\nabla_\alpha f(r)=f^{'}(r)\nabla_\alpha r$.  For the
bare Coulomb potential, $f(r)=1/r$, and thus 
$\nabla_\alpha (1/r)=-(1/r^2)\nabla_\alpha r\equiv T_\alpha^{0}$. Combining these two simple results the gradient of a general
function $f(r)$ may be expressed in terms of the bare Coulomb tensor 
via $\nabla_\alpha f(r)=-r^2 f^{'}(r)T_\alpha^0$. Introducing the 
differential operator ${\cal D}$, defined by  ${\cal D}f(r) =-r^2 f^{'}(r)$, 
this may furthermore be written as $\nabla_\alpha f(r) = {\cal D}f(r) T_\alpha^0$.  Thus the set of successive interaction tensors for a screened Coulomb Potential $T^{\kappa}$ may be written, using the product rule of
differentiation  as:
\begin{eqnarray}
\label{tensorkappa}
T_\alpha^\kappa&=&\nabla_\alpha f(r)=({\cal D}^1f) T_\alpha^{0}\nonumber\\
T_{\alpha\beta}^{\kappa} &=& \nabla_\alpha \nabla_\beta f(r) =({\cal D}^2 f)T_\alpha^0T_\beta^0 + ({\cal D}^1 f) T_{\alpha\beta}^0\nonumber\\
T_{\alpha\beta\gamma}^\kappa&=& ({\cal D}^3f)T_\alpha^0 T_\beta^0T_\gamma^0 +
({\cal D}^2 f)[T_{\alpha\beta}^0T_\gamma^0 + T_{\alpha\gamma}^0 T_\beta^0+T_{\beta\gamma}^0 T_\alpha^0]+({\cal D}^1f)T_{\alpha\beta\gamma}^0\nonumber\\
T_{\alpha\beta\gamma\delta}^{\kappa}&=& ({\cal D}^4f)T_\alpha^0 T_\beta^0 T_\gamma^0 T_\delta^0\nonumber\\
 &+& ({\cal D}^3f)[T_{\alpha\beta}^0T_\gamma^0T_\delta^0 + T_{\beta\gamma}^0T_\alpha^0T_\delta^0 + T_{\alpha\delta}^0 T_\beta^0 T_\gamma^0 +
T_{\beta\delta}^0 T_{\alpha}^0 T_\gamma^0 + T_{\alpha\gamma}^0 T_\beta^0 T_\delta^0 +T_{\gamma\delta}^0 T_\alpha^0 T_\beta^0] \nonumber\\
&+&({\cal D}^2f)[T_{\alpha\beta\gamma}^0 T_\delta^0 + T_{\alpha\beta\delta}^0 T_\gamma^0 + T_{\alpha\gamma\delta}^0T_{\beta}^0 +T_{\beta\gamma\delta}^0 T_\alpha^0 + T_{\alpha\beta}^0T_{\gamma\delta}^0 + T_{\alpha\gamma}^0 T_{\beta\delta}^0 + T_{\alpha\delta}^0T_{\beta\gamma}^0] \nonumber\\
&+& ({\cal D}^1f)T_{\alpha\beta\gamma\delta}^0,
\end{eqnarray}
where the coefficients $\{{\cal D}^nf\}$ are functions solely of distance, given by
\begin{eqnarray}
\label{Dflist}
{\cal D}^1f&=&(1+\kappa r) e^{-\kappa r}\nonumber\\
{\cal D}^2f&=&\kappa^2 r^3 e^{-\kappa r}\nonumber\\
{\cal D}^3f&=&\kappa^2 r^4(\kappa r-3)e^{-\kappa r}\nonumber\\
{\cal D}^4f&=&\kappa^2 r^5(12-8\kappa r +\kappa^2 r^2)e^{-\kappa r},
\end{eqnarray}
and all factors of $1/4\pi\epsilon_0\epsilon$ and more significantly all
angular dependencies of the interaction are embodied in the
bare Coulomb tensors $T^0_{\alpha\beta\cdots}$, expressions for which
are easily calculated.

\section{Evaluation of Interaction Energies}

The interaction energy between two discs (A and B) , separated by a distance
$r_{AB}$ and oriented at spherical polar angles $(\theta_A,\phi_A)$ and $(\theta_B,\phi_B)$ may be written, eq.(\ref{utottot}), as a sum of contributions to the
total from the interactions of each order of multipole via
\begin{equation}
\label{app:Uabtot}
V_{AB}(r_{AB},\theta_A,\theta_B,\phi_A,\phi_B)=V_{AB}^{ZZ}+ (V_{AB}^{ZQ}+V_{AB}^{QZ}) + V_{AB}^{QQ} + (V_{AB}^{Z\Phi}+V_{AB}^{\Phi Z}) + \cdots,
\end{equation}
where $Z,Q$ and $\Phi$ denote respectively the charge, quadrupole moment and
hexadecapole moment on a disc, and $V_{AB}^{QZ}$ for instance denotes the
contribution from the interaction of the charge on disc $A$ with the quadrupole
on disc $B$.

The leading term in this expansion is simply the screened Coulomb interaction 
of the two charges, given by 
\begin{equation}
\label{app:Uabzz}
V_{AB}^{ZZ}= Z^Ae T^{\kappa} Z^Be = \frac{Z^2e^2}{4\pi\epsilon_0\epsilon} \frac{e^{-\kappa r}}{r}
\end{equation}
where $T^{\kappa}$ is the zeroth-order interaction tensor.  The next term in eq.(\ref{app:Uabtot}) corresponds to the charge-quadrupole interaction, $V_{AB}^{ZQ}$,
which is written in the following form
\begin{equation}
\label{app:Uabqz}
V_{AB}^{ZQ}= \frac{e^2}{2!}Z^A T_{\alpha\beta}^{\kappa} Q^B_{\alpha\beta}
\end{equation}
Expressing the screened interaction tensor $T_{\alpha\beta}^{\kappa}$ in terms of the unscreened tensors $\{T_{\alpha\beta\cdots}^0\}$, using eq.(\ref{tensorkappa}), and recalling  the definition of the 
quadrupole moment tensor, eq.(\ref{defineq}) this sum is calculated via:
\begin{eqnarray}
\label{QabTabsum}
T_{\alpha\beta}^\kappa Q_{\alpha\beta}^B &=& -Q[{\cal D}^2f
 T_\alpha^0 T_\beta^0 + {\cal D}^1 f T_{\alpha\beta}^0][\delta_{\alpha\beta}-n_\alpha^B n_\beta^B] \nonumber \\
&=& -Q\left\{{\cal D}^2f[T_\alpha^0T_\alpha^0- T_\alpha^0 n_\alpha^B T_\beta^0 n_\beta^B] +{\cal D}^1 f[T_{\alpha\alpha}^0 -n_\alpha^B T_{\alpha\beta}^0 n_\beta^B]\right\}\nonumber\\
&=& -\frac{Q}{4\pi\epsilon_0\epsilon} \left[{\cal D}^2 f \left(\frac{1-\cos^2\theta_B}{r^4}\right) +{\cal D}^1 f \left(\frac{1-3\cos^2\theta_B}{r^3}\right)\right] \\
&=& \frac{Za^2}{4}\frac{1}{4\pi\epsilon_0\epsilon}\frac{e^{-\kappa r}}{r^3}\left[(1+\kappa r +\kappa^2 r^2) -\cos^2\theta_B(3 +3\kappa r + \kappa^2 r^2)\right] \nonumber,
\end{eqnarray}
where $\{n_\alpha^B\}$ denote the Cartesian components of the  unit vector 
which defines the 
major axis of disc $B$, and the $\{{\cal D}^nf\}$ are defined by eq.(\ref{Dflist}). Along with this energy, the contribution due to the interaction between quadrupole on disc A and charge on disc B must be added.
When all multiplicative factors have been included the final result reads
\begin{eqnarray}
\label{app:uabqzfinal}
V_{AB}^{ZQ+QZ}&\equiv& V_{AB}^{ZQ+QZ}(r,\theta_A,\theta_B)\nonumber\\&=&-\frac{Z^2e^2a^2}{8}\left(\frac{1}{4\pi\epsilon_0\epsilon}\right)\frac{e^{-\kappa r}}{r^3}\times\\
&\ &\left[\left(1+\kappa r +\frac{\kappa^2 r^2}{3}\right)\left(3\cos^2\theta_A+3\cos^2\theta_B\right) -2\left(1+\kappa r +\kappa^2 r^2\right)\right],\nonumber
\end{eqnarray}
\noindent where it is observed that the charge-quadrupole interaction has
no dependence on the azimuthal angles $\phi_A$ and $\phi_B$.

In the quadrupole-quadrupole interaction energy it is  necessary to calculate  the sum
$Q_{\alpha\beta}^A T_{\alpha\beta\gamma\delta}^\kappa Q_{\gamma\delta}^B$. The
screened tensor is expressed as a sum of terms involving the simpler  unscreened tensors  using eq.(\ref{tensorkappa}), and simplified
further by expressing all second and higher rank unscreened tensors in terms
of the first order unscreened tensors, via:
\begin{eqnarray}
\label{tn0lots}
4\pi\epsilon_0\epsilon T_\alpha^0 &=& -\frac{r_\alpha}{r^3} \nonumber \\
4\pi\epsilon_0\epsilon T_{\alpha\beta}^0 &=& \frac{3r_\alpha r_\beta-r^2\delta_{\alpha\beta}}{r^5}\equiv r[3 T_\alpha^0 T_\beta^0 - \frac{\delta_{\alpha\beta}}{r^4}]\nonumber\\
4\pi\epsilon_0\epsilon T_{\alpha\beta\gamma}^0 &\equiv& 15rT_\alpha^0T_\beta^0T_\gamma^0 -\frac{3}{r^2}(T_\alpha^0\delta_{\beta\gamma} +  T_{\beta}^0\delta_{\alpha\gamma}+T_\gamma^0\delta_{\alpha\beta}).
\end{eqnarray}
Following this procedure of expressing the elements of the 
$n^{th}$-rank unscreened tensors
in terms of those of the  $1^{st}$ rank tensors, the    
full unscreened $4^{th}$ rank tensor, using (\ref{tensorkappa}) and(\ref{tn0lots}), is given by
\begin{eqnarray}
\label{app:4tk}
T_{\alpha\beta\gamma\delta}^{\kappa} &=& 
T_\alpha^0 T_\beta^0 T_\gamma^0 T_\delta^0[{\cal D}^4 f +18r {\cal D}^3 f +87 r^2 {\cal D}^2 f] \nonumber\\
&+&[ T_\alpha^0T_\beta^0\delta_{\gamma\delta} +  T_\alpha^0T_\gamma^0\delta_{\beta\delta}+T_\alpha^0T_\delta^0\delta_{\beta\gamma} + T_\beta^0T_\gamma^0\delta_{\alpha\delta} +T_\beta^0T_\delta^0\delta_{\alpha\gamma} + T_\gamma^0T_\delta^0\delta_{\alpha\beta}]\nonumber\\
&\ & [-({\cal D}^3 f/r^3) -9({\cal D}^2f/r^2)]\nonumber\\
&+& [\delta_{\alpha\beta}\delta_{\gamma\delta} +\delta_{\alpha\gamma}\delta_{\beta\delta}+\delta_{\alpha\delta}\delta_{\beta\gamma}][{\cal D}^2f/r^6]\nonumber\\
&+& {\cal D}f T_{\alpha\beta\gamma\delta}^0
\end{eqnarray}
where the tensor $T_{\alpha\beta\gamma\delta}^0$ appearing in the last line of eq.(\ref{app:4tk}) is the only  term surviving if $\kappa=0$, corresponding to the purely Coulombic interaction.   

The screened Coulombic tensor $T_{\alpha\beta\gamma\delta}^\kappa$, given by eq.(\ref{app:4tk}),  is used to calculate both the contribution to the potential away from a single disc due to the hexadecapole moment $\Psi^{\Phi}({\pmb r})$ and
also the quadrupole-quadrupole and charge-hexadecapole energies in the
effective pair potential. For illustrative purposes the
quadrupole-quadrupole interaction energy shall be pursued here, 
which involves the sum $Q_{\alpha\beta}^{A}T_{\alpha\beta\gamma\delta}^{\kappa} Q_{\gamma\delta}^{B}$. Using eq.(\ref{app:4tk}) it is evident that this sum will itself be the sum of contributions
from terms involving $Q_{\alpha\beta}^AT_\alpha^0T_\beta^0T_\gamma^0T_\delta^0 Q_{\gamma\delta}^B$, $Q_{\alpha\beta}^AT_\alpha^0T_\beta^0\delta_{\gamma\delta}Q_{\gamma\delta}^B$ etc. which must each be calculated separately.  The first of these is calculated as
\begin{eqnarray}
\label{app:qqtemp}
Q_{\alpha\beta}^A T_\alpha^0T_\beta^0T_\gamma^0T_\delta^0 Q_{\gamma\delta}^B &=& Q^2[\delta_{\alpha\beta}-n_\alpha^An_\beta^A][T_\alpha^0T_\beta^0T_\gamma^0T_\delta^0][\delta_{\gamma\delta} - n_\gamma^Bn_\delta^B]\nonumber\\
&=& Q^2[T_\alpha^0T_\alpha^0 - T_\alpha^0 n_\alpha^A T_\beta^0 n_\beta^A][T_\gamma^0T_\gamma^0 - T_\gamma^0n_\gamma^BT_\delta^0n_\delta^B]\nonumber\\
&=& \frac{Q^2}{4\pi\epsilon_0\epsilon}\frac{(1-\cos^2\theta_A)(1-\cos^2\theta_B)}{r^8}
\end{eqnarray}
Proceeding along these lines for each of the terms appearing in $Q_{\alpha\beta}^{A}T_{\alpha\beta\gamma\delta}^{\kappa} Q_{\gamma\delta}^{B}$ the interaction
energy finally reads:
\begin{eqnarray}
\label{qatkqb}
V_{AB}^{QQ}&\equiv& \frac{e^2}{2!2!}Q_{\alpha\beta}^A T_{\alpha\beta\gamma\delta}^\kappa Q_{\gamma\delta}^B\nonumber\\
&=& \frac{Z^2e^2 a^4}{64}(\delta_{\alpha\beta}-n_\alpha^A n_\beta^A) T_{\alpha\beta\gamma\delta}^\kappa (\delta_{\gamma\delta}-n_\gamma^B n_\delta^B)\nonumber\\
&=&\frac{Z^2e^2 a^4}{64}\left(\frac{1}{4\pi\epsilon_0\epsilon}\right)\left\{ [{\cal D}^4 f +18r {\cal D}^3f +87r^2 {\cal D}^2 f]\frac{1}{r^8}(1-\cos^2\theta_A)(1-\cos^2\theta_B) \right.\nonumber \\
&\ & \hspace{0.5cm} -[{\cal D}^3f +9r{\cal D}^2f]\frac{1}{r^7}[8-6\cos^2\theta_A -6\cos^2\theta_B \nonumber\\ 
&\ & \hspace{1.5cm} +4\cos\theta_A\cos\theta_B(\cos\theta_A\cos\theta_B +\sin\theta_A\sin\theta_B\cos(\Delta\phi))]\nonumber\\
&\ & \hspace{0.5cm} +\frac{{\cal D}^2f}{r^6}[6+2(\cos\theta_A\cos\theta_B +\sin\theta_A\sin\theta_B\cos(\Delta\phi))^2]\nonumber \\
&\ & \hspace{0.5cm} +  \frac{{\cal D}f}{r^5}[3-15\cos^2\theta_A -15\cos^2\theta_B -45\cos^2\theta_A\cos^2\theta_B\nonumber\\
&\ & \hspace{1.5cm} \left. +6(4\cos\theta_A\cos\theta_B -\sin\theta_A\sin\theta_B\cos(\Delta\phi))^2]\right\}
\end{eqnarray}
where attention may be drawn to the fact that the interaction energy
only involves the difference in the azimuthal angles, $\Delta\phi=\phi_A-\phi_B$ and not on their absolute values.
\end{appendix}

\bibliography{references}

\newpage
\section*{Table Captions}
\begin{enumerate}
\item
\label{table1}
Coefficients appearing in the series expansion of the electrostatic potential 
along the $z-$axis, eq.(\ref{screen-z}). $C_{\frac{1}{2}}^n$ is the 
coefficient of the term of order $x^n$ in the binomial expansion of 
$(1+x)^{\frac{1}{2}}$.
\end{enumerate}
\section*{Figure Captions}
\begin{enumerate}
\item
\label{figure1} 
Integration over the surface of a disc
\item
\label{figure2}
Electrostatic potential along the z-axis. Solid lines denote the
multipolar expansion, dashed lines the linearized PB potential, and triangles
a numerical integration over a discretized charge distribution (Yukawa segment model).  The upper set of curves are for $\kappa a=0.5$, and the lower for 
$\kappa a=1.0$.  The divergence is highlighted in the logarithmic inset figure.
\item
\label{figure3}
Electrostatic potential in the $xy$-plane, symbols as in fig.(\ref{figure2}) for
$\kappa a=0.5$,$1.0$.
\item
\label{figure4}
Geometry of a pair of platelets. The disc orientations are 
characterized by the spherical polar angles $\theta$ and $\phi$. 
The azimuthal angle 
$\phi_B$ has been omitted for clarity.
\item
\label{figure5}
Contributions to the Potential Energy as a function of
separation for two parallel coaxial plates,($\theta_A=0, \theta_B=0,\Delta\phi=0$),  for inverse screening length $\kappa =0.5/a$. The points plotted on the LPB curve (known in this geometry)  correspond to a discretized Yukawa Segment model calculation.

\item
\label{figure6}
Contributions to the Potential Energy as a function of
separation for two parallel coplanar plates, ($\theta_A=\pi/2, \theta_B=\pi/2,\Delta\phi=0$), again for $\kappa a=0.5$. The
discretized Yukawa segment calculation has been included for comparison.

\item
\label{figure7}
Contributions to the Potential Energy as a function of
separation for two plates in T-shaped configuration,($\theta_A=0, \theta_B=\pi/2,\Delta\phi=0$) with $\kappa a=0.5$. The discretized
Yukawa segment calculation has been included for comparison.

\item
\label{figure8}
Contributions to the Potential Energy as a function of
the horizontal separation, r/a, for parallel plates with plate $B$ fixed at a
altitude of $1.5a$ above disc $A$ and $\kappa a=0.5$. The discretized Yukawa
segment calculation has been included for comparison.

\item
\label{figure9}
Contributions to the Potential Energy as a function of the
angle $\theta_B$ with $\theta_A=0$ for  $\kappa a =0.5$, and fixed
centre-to-centre distance $r=1.5a$.  The discretized Yukawa segment
calculation has been included for comparison.

\item
\label{figure10}
Dependence of the energy on the difference in 
azimuthal angle $\Delta \phi$, plotted 
for fixed $\theta_A=\theta_B=\pi/4$ for disc separations 
$r/a=$ 1.0,1.1,1.2,1.3,1.4 and 1.5 from top to bottom, for 
$\kappa a=0.5$.
\end{enumerate}
\newpage

\begin{center}
\begin{tabular}[h]{|c|c|c|}
\hline
n & $A_n$ & $K_n$ \\
\hline
$0$ & $+1$ & $0$ \\ & & \\
$1$ & $0$ & $-\frac{\kappa a^2}{4}$\\ & & \\
$2$ & $-\frac{a^2}{4}$ & $+\frac{\kappa^2 a^4}{24}$ \\ & &  \\
$3$ & $0$ & $+\frac{\kappa a^4}{8}-\frac{\kappa^3 a^6}{192}$\\ & &  \\
$4$ & $+\frac{a^4}{8}$ & $-\frac{\kappa^2 a^6}{32} + \frac{\kappa^4 a^8}{1920}$\\ & &  \\
$5$ & $0$ & $-\frac{5\kappa a^6}{64} +\frac{\kappa^3 a^8}{192} -\frac{\kappa^5 a^{10}}{23040}$ \\
& & \\
\vdots & \vdots&\vdots \\
\hline
$n\ {\rm  even}$ & $ 2 a^nC_{\frac{1}{2}}^{(n/2)+1}$ & \vdots\\
$n\ {\rm odd}$ & $0$ & \vdots\\
\hline
\end{tabular}
\center{Table 1}
\end{center}
\vspace{1.0cm}
\begin{figure}[h]
\begin{center}
\epsfig{file=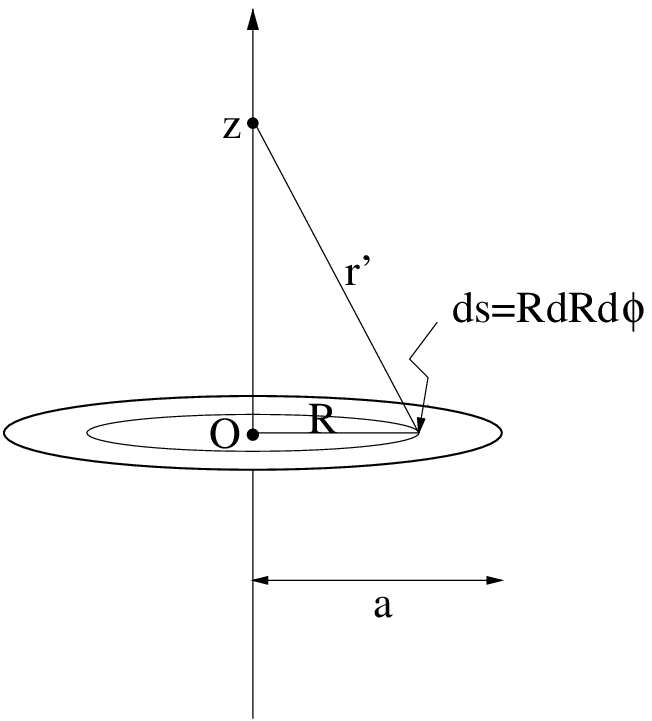}
\center{Figure 1}
\end{center}
\end{figure}
\newpage

\begin{figure}[h]
\begin{center}
\epsfig{file=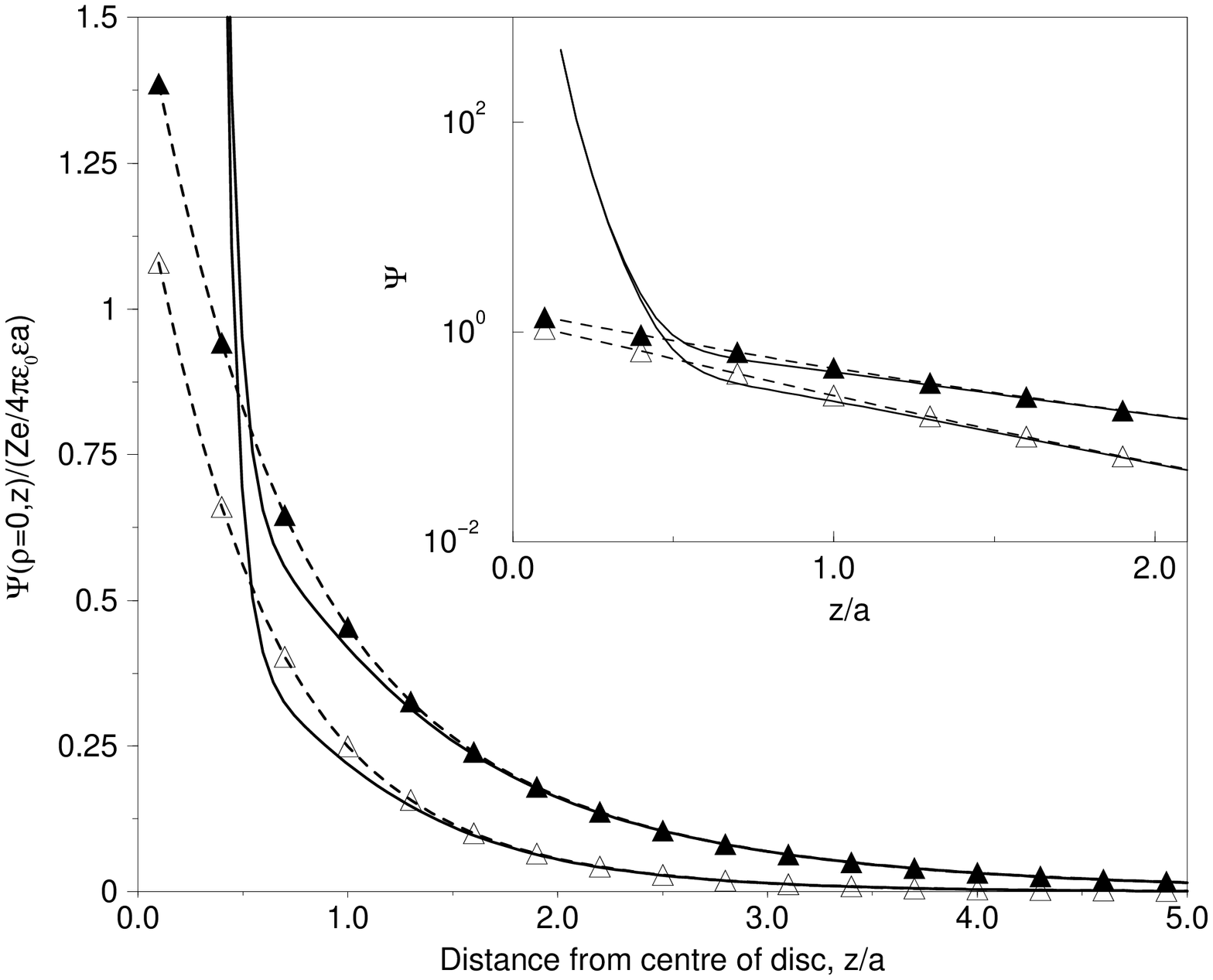,height=14.0cm,angle=270}
\center{Figure 2}
\end{center}
\end{figure}
\newpage

\begin{figure}[h]
\begin{center}
\epsfig{file=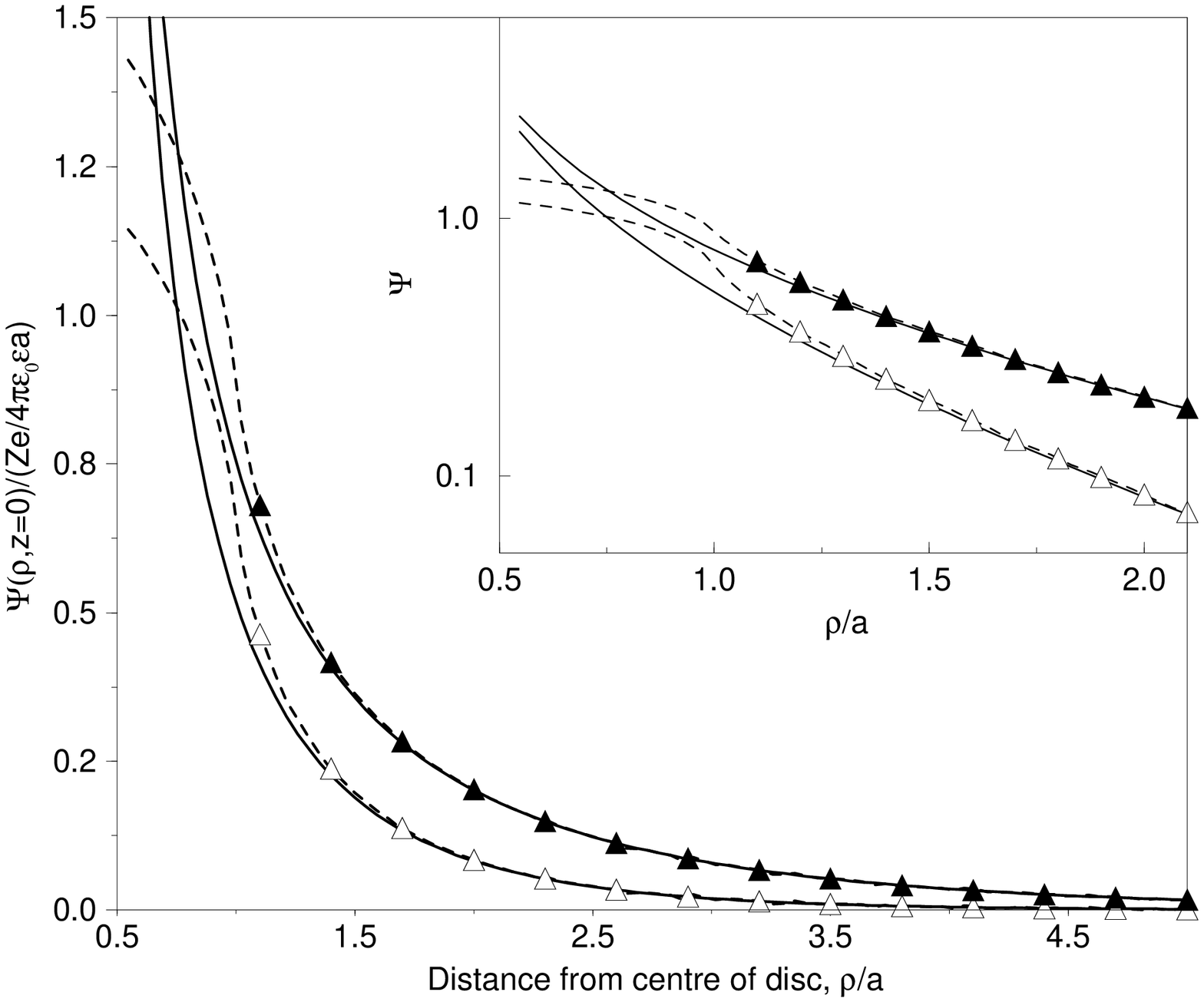,height=14.0cm,angle=270}
\center{Figure 3}
\end{center}
\end{figure}
\newpage

\begin{figure}[h]
\begin{center}
\epsfig{file=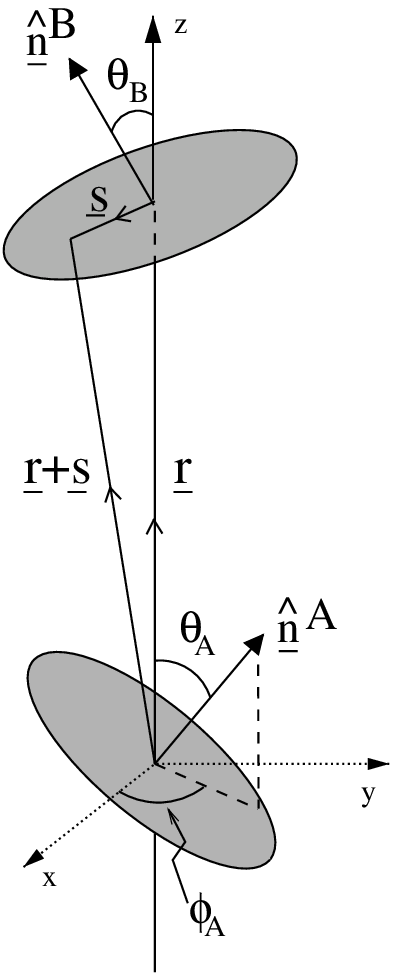}
\center{Figure 4}
\end{center}
\end{figure}
\newpage

\begin{figure}[h]
\begin{center}
\epsfig{file=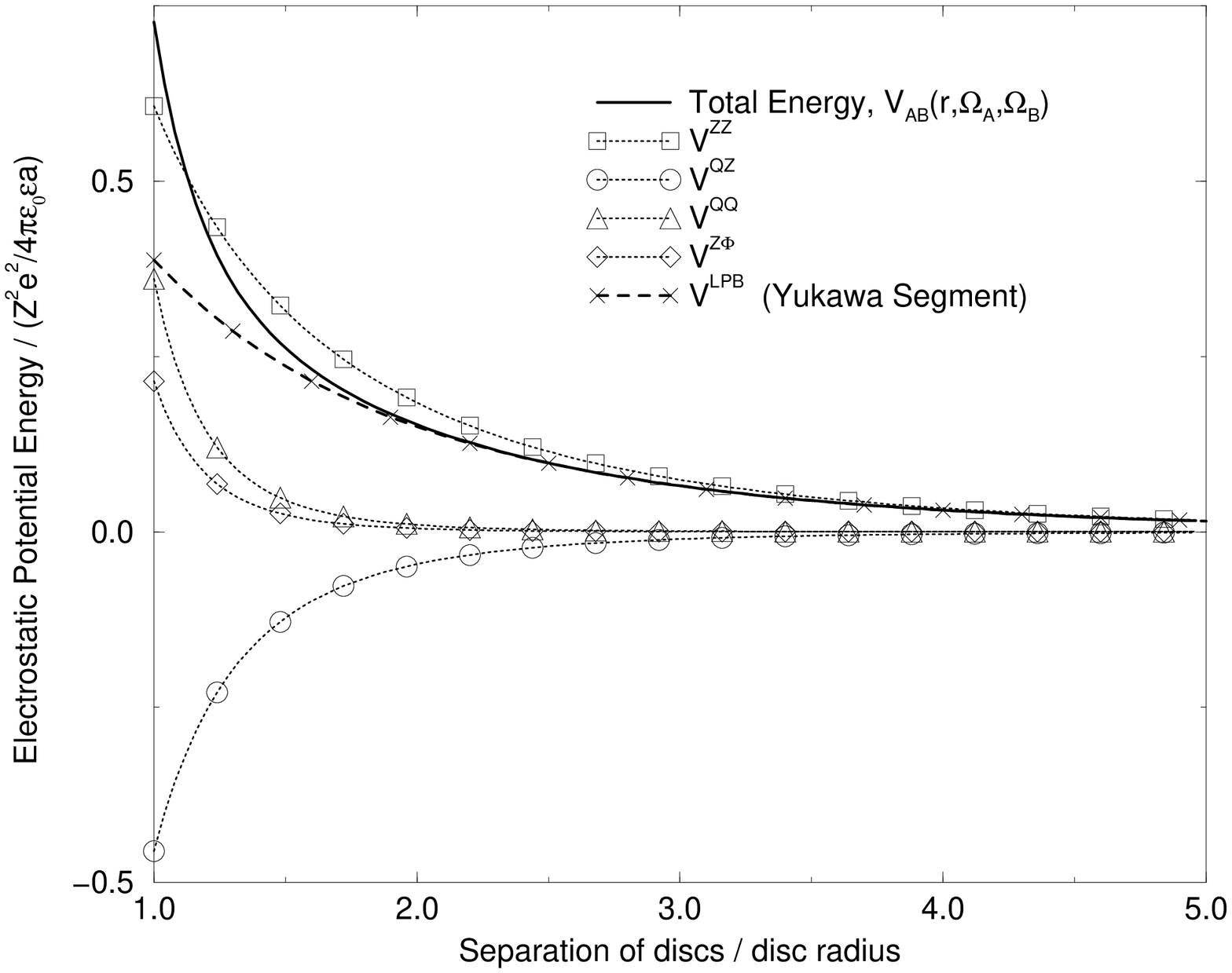,height=15cm,angle=270}
\center{Figure 5}
\end{center}
\end{figure}
\newpage

\begin{figure}[h]
\begin{center}
\epsfig{file=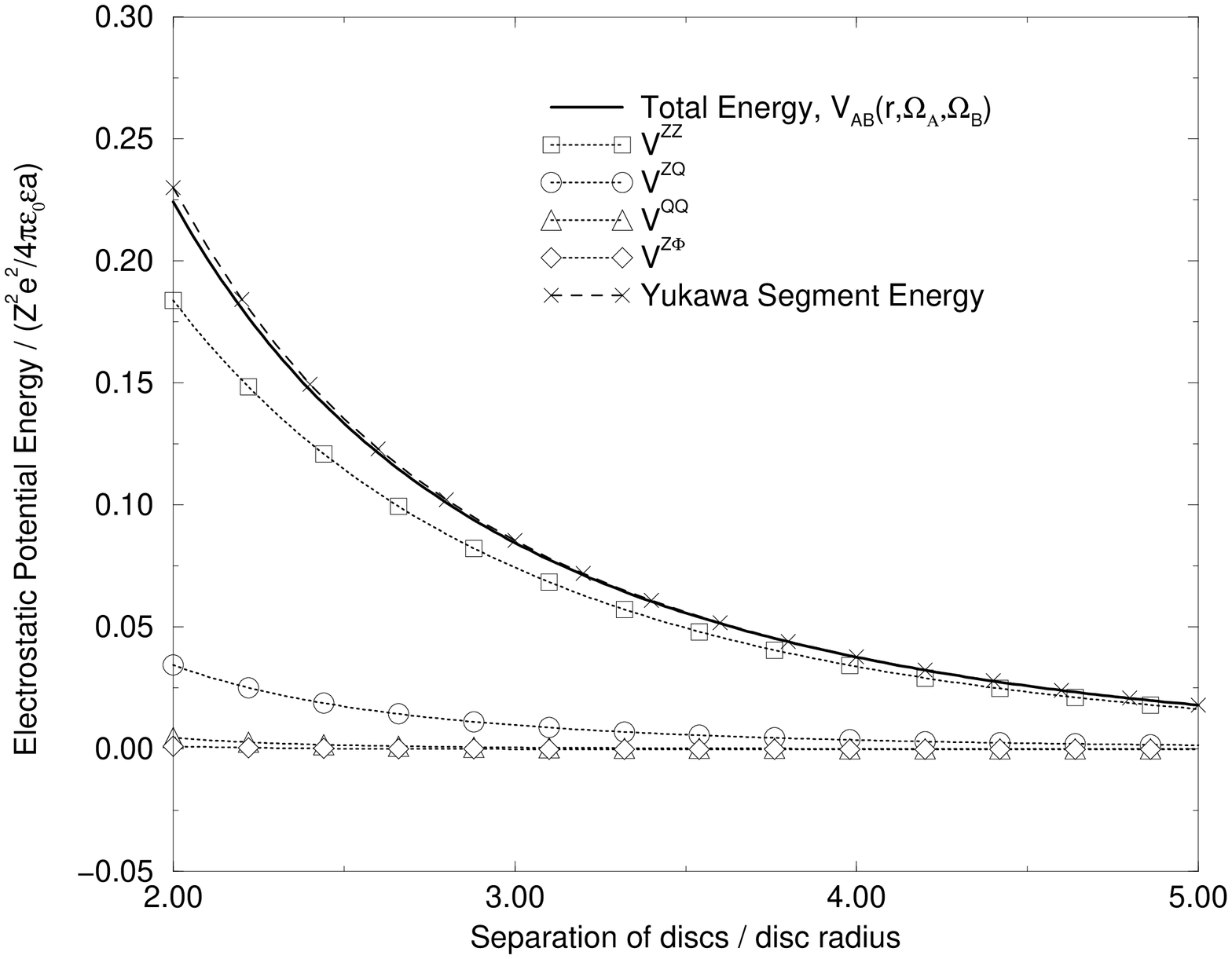,height=15cm,angle=270}
\center{Figure 6}
\end{center}
\end{figure}
\newpage

\begin{figure}[h]
\begin{center}
\epsfig{file=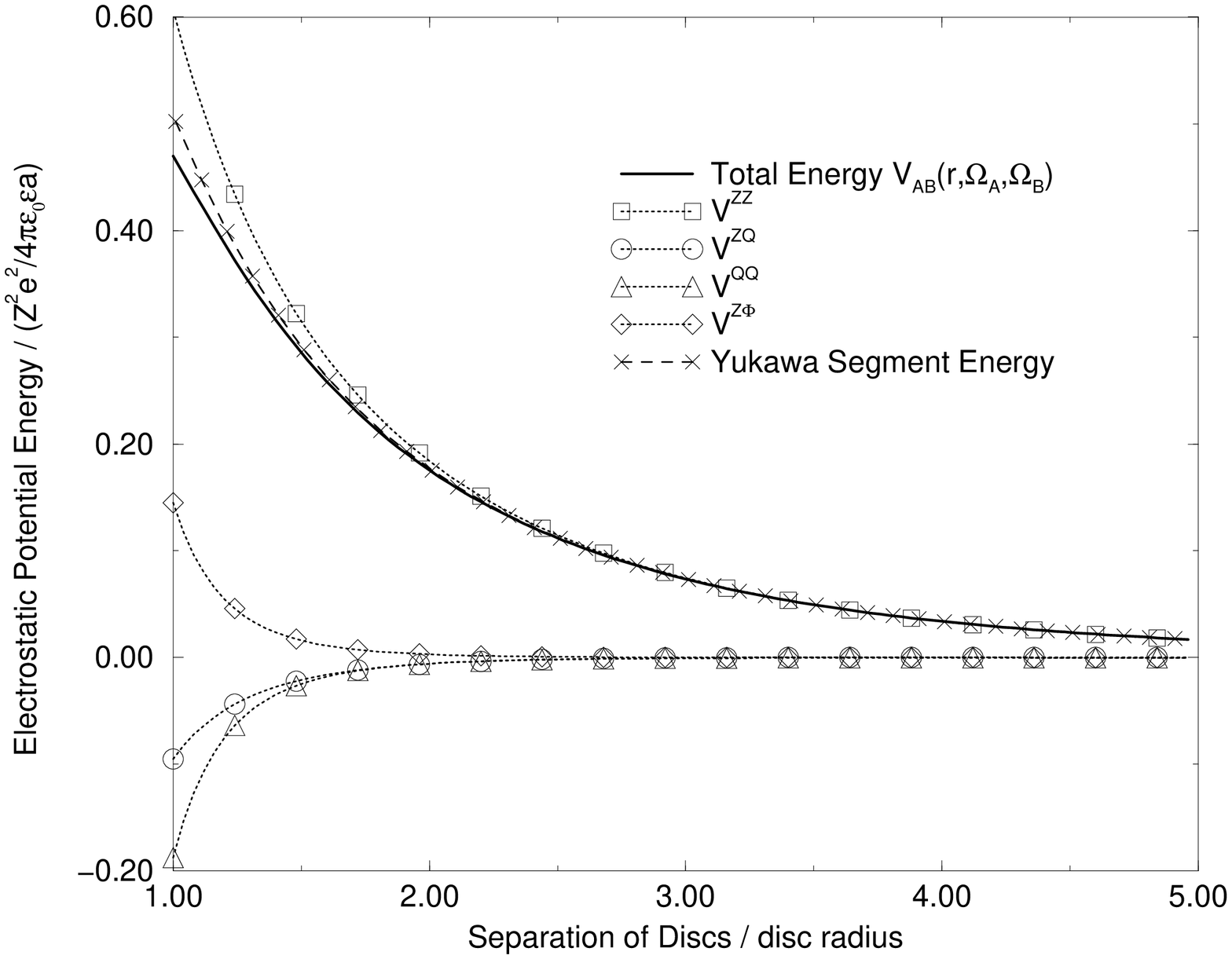,height=15cm,angle=270}
\center{Figure 7}
\end{center}
\end{figure}
\newpage

\begin{figure}[h]
\begin{center}
\epsfig{file=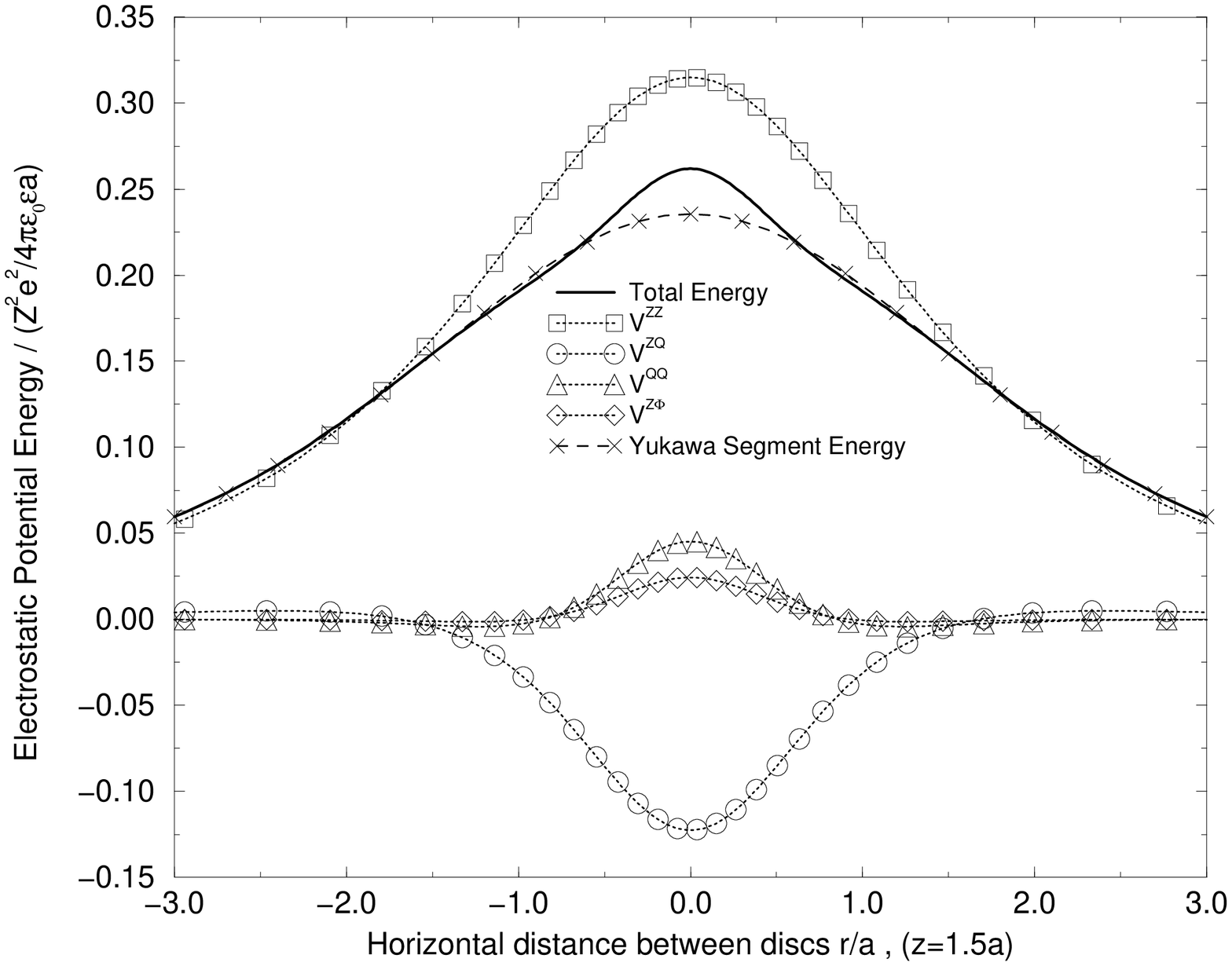,height=15cm,angle=270}
\center{Figure 8}
\end{center}
\end{figure}
\newpage

\begin{figure}[h]
\begin{center}
\epsfig{file=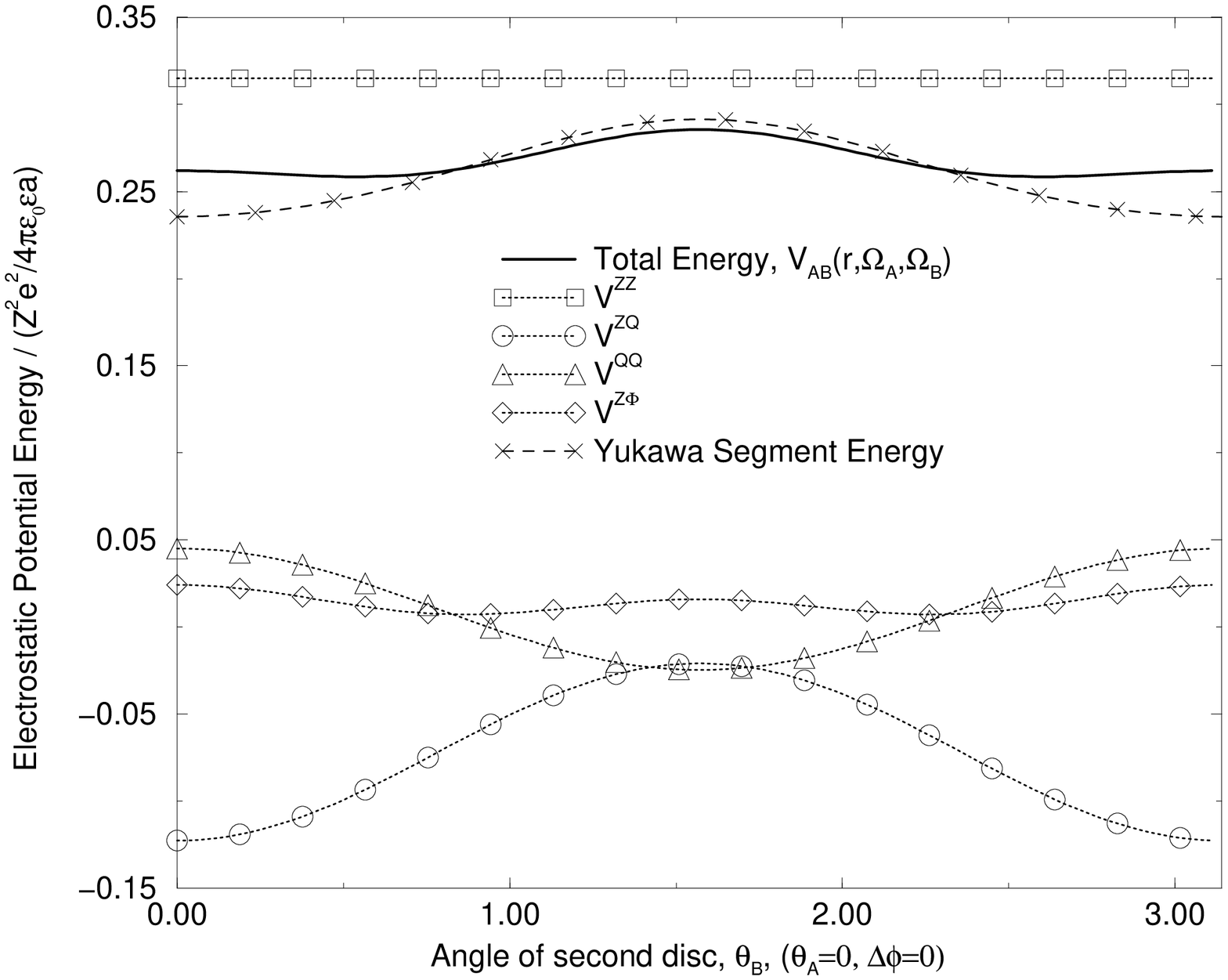,height=15cm,angle=270}
\center{Figure 9}
\end{center}
\end{figure}

\begin{figure}[h]
\begin{center}
\epsfig{file=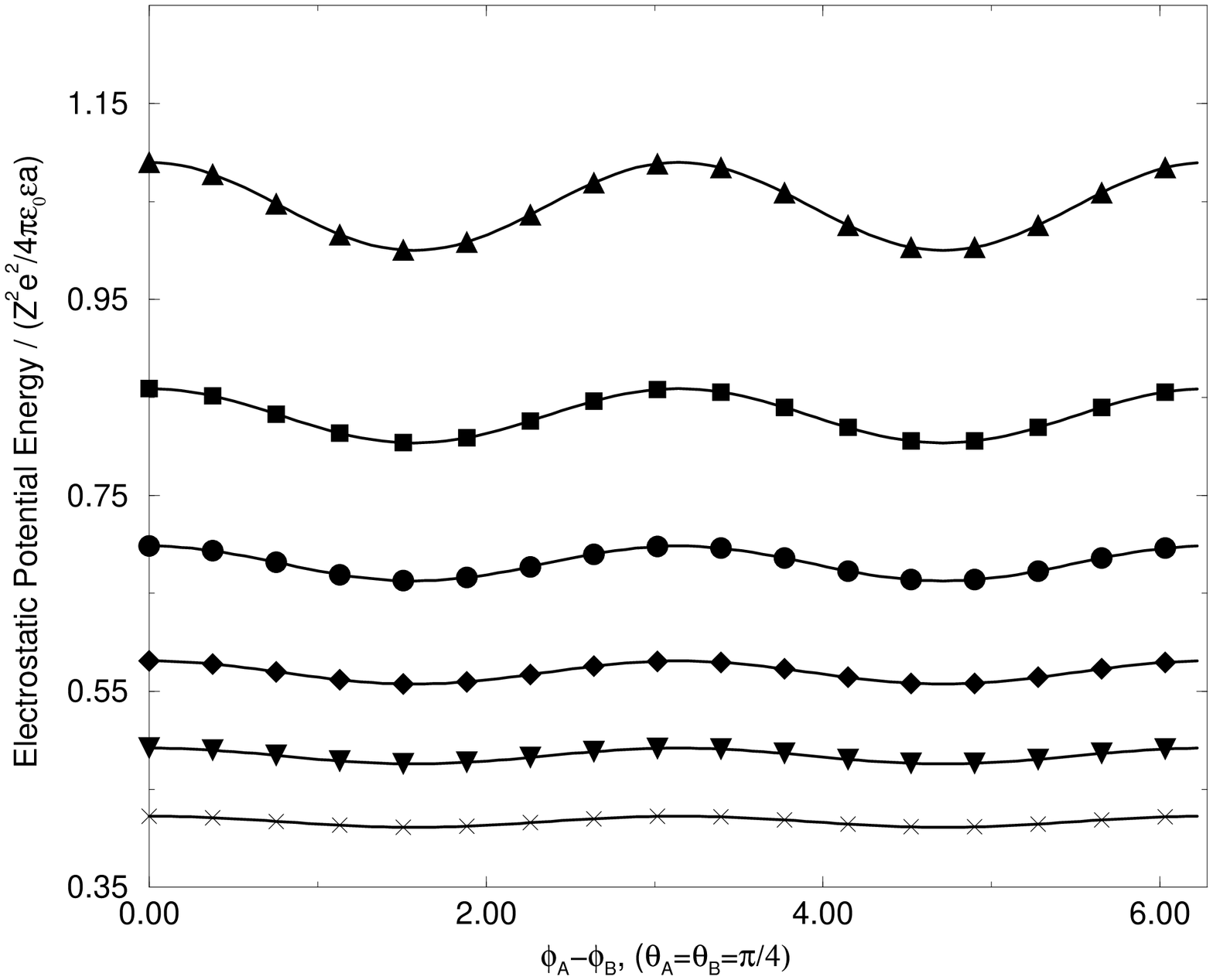,height=15cm,angle=270}
\center{Figure 10}
\end{center}
\end{figure}

\end{document}